\documentstyle[preprint,aps]{revtex}
\begin{document}
\title{Direction dependency of extraordinary refraction index}
\author{Mojca \v Cepi\v c}
\address{Faculty of Education, University of Ljubljana, Kardeljeva
pl.16, 1000 Ljubljana, Slovenia \\
Jo\v zef Stefan Institute, Jamova 39, 1000 Ljubljana, Slovenia }
\date{\today}
\maketitle
\begin{abstract}
A simple experiment is presented that enables qualitative and quantitative
measurement of the extraordinary refractive index direction
dependency in an uniaxial nematic liquid crystal. Three liquid crystaline
cells were designed in which elongated molecules of nematic liquid
crystal align in directions which enable to observe the variation
of extraordinary refractive index as a function of the direction of
light.
\end{abstract}
\newpage
\tightenlines
The most common experiment to demonstrate birefringence is the
observation of doubled objects through the calcite crystal$^1$.
When a polarizer is placed behind the crystal or in front of
it, one of the pictures dissapears if the polarizer
direction coincides with one of the polarizability tensor
eigenvectors. Although the doubling of pictures is persuasive
for an educated physicist, an extensive explanation is needed for
students to whom the phenomenon is presented for the first time.

Another experiment which demonstrates the splitting of the light
beam into two perpendicularly polarized beams with different phase
velocities uses a large birefringent crystal, thick enough that
after the transmition of the incident nonpolarized beam, two
separated beams can be observed as two light spots on a distant
screen. Changing the direction of the incident light by rotating
the sample, the direction dependency of the extraordinary
refraction index in uniaxial crystals as well as the direction
dependency of both indeces in biaxial crystals can be observed.
Unfortunately, the accuracy of the quantitative
measurements of both indeces is poor, since only a slight nonparallelism
of the sample surfaces can results in huge changes of the transmitted
light direction$^2$. The easiest accesible crystals
are biaxial ( e.g. quarz), which additionally complicates the
comparison of the results with theoretical predictions.

In this paper a simple experiment is presented in which liquid crystal in the optically
uniaxial nematic phase is used as a
birefringent material. The experiment enables demonstration of
the extraordinary refraction index direction dependency. In the
undergraduate lab the experiment can also be used for accurate
measurements
of the extraordinary index in different directions. The
characteristic experimental data is presented in the present paper.

Nematic liquid crystals are composed of elongated molecules with
orientationally ordered long molecular axes. Due to the rapid molecular
rotations around long molecular axis, the system is optically uniaxial.
The orientational correlation length perpendicular to the molecular long
 axes  extends to 500 $\mu$m in nematic liquid crystals$^3$,
therefore the well ordered samples have to be thinner than that. Since
elongated molecules have a larger polarizability along the long
molecular axis than perpendicular to it, the direction of the
optical axis for a liquid crystal in a cell is known from the
preparation of the glass coating. In thin samples, spatial
separation of the ordinary and extraordinary beam can be obtained
by the prismatic effect$^4$.

In order to study the direction dependency of the extraordinary
refraction index, three different wedge cells shown in Fig.1 were
designed. The cells were approximately 1 cm long and half of the
cm wide. The 200 $\mu$m thick foil was inserted and glued between
the glasses in one of the narrower sides, while the other narrow
side of the cell was glued directly together forming the wedge.
Two of the cells had polymide coating rubbed along the wedge of
the cell or perpendicular to it (Fig. 1 a,b). The third cell had a
polymer coating which aligned long molecular axes perpendicular to
the glass (Fig. 1c). To avoid a long description of molecular
orientations in different cells, let us call the cell with
molecular axes oriented along the wedge (Fig. 1a) the {\it
transverse cell}, the cell  with long axes oriented along the
longer side of the cell (Fig. 1b) the {\it longitudinal cell} and
the cell with molecular long axes perpendicular to the glass (Fig.
1c) the {\it perpendicular cell}. The names should remind the
reader of the orientations of the long molecular axes and
consequently of the direction of the extraordinary polarization
and the optical axes. The cells were filled with the liquid
crystal$^5$ heated above the transition temperature from the nematic
to the isotropic  phase. The capilary effect was used to fill the
cell.

The phase velocity of light in the direction in an angle $\theta$
with the optical axis of an uniaxial crystal like a nematic liquid
crystal is given by$^6$
\begin{equation}
v_e(\theta)^2 = v_o^2 \cos^2 \theta + v_{e,0}^2 \sin^2 \theta.
\label{start}
\end{equation}
Here $v_e$~ is the phase velocity of light with the extraordinary
polarization i.e. with the electric field oscillating in the
incident plane. The phase velocity $v_{e,o}$~ is the smallest
phase velocity of the extraordinary polarized light in a negative
birefringent uniaxial liquid crystal, when the light direction is
perpendicular to the optical axis i.e. $\theta = 90^\circ$. The
ordinary phase velocity $v_o$~ is direction independent. Both
refractive indeces can be obtained from the general definition of
a refractive index as a ratio between the phase velocity of light
in vacuum and the phase velocity in a transparent material:
\begin{equation}
n_e(\theta)^2 = {{n_{e,0}^2\;n_o^2}\over{n_{e,0}^2\;\cos^2 \theta
+n_o^2\; \sin^2 \theta}}
\label{index}
\end{equation}
where refraction indeces are marked with the same subscripts as the
corresponding phase velocities. The light polarized in the incident plane
therefore refracts differently that the light polarized
perpendicular to the incident plane. The direction of the extraordinary
light in the birefringent material therefore depends on the orientation
of the optical axis as well as the incident angle.

The experimental setup is shown in Fig. 2. The wedge cell
is fixed into the holder and it is put on the rotatable table with
longer side parallel to the table surface. A laser pointer is used
as a source of the
nonpolarized light beam. In addition to the low price, laser pointer has
another advantage over the He-Ne lasers, i.e. it is a nonpolarized
quasi monocromatic light source. The
dispersion of both refractive indeces is quite common in liquid crystals
and the experimental study with a monochromatic light avoids
this problem. The direction of the incident light is always in
the incident plane perpendicular to the wedge. When passing through the wedge
cell the
nonpolarized incident beam splits into two perpendicularly
polarized beams. After the transmition both beams have different directions
due to the prismatic effect of
the wedge and after a few tenths of a meter they become separated in
space. On a distant screen two light dots appear (Fig. 2). We measure the
position of the dots relative to the position of the direct beam
dot, when light does not pass through the cell.
Changing the incident direction of light by rotating the
table, the position of dots changes and enable the calculation of
both indeces.

In the transverse cell none of the indeces
changes with the direction of light as long as the incident
plane is parallel to the long side of the cell (Fig. 1a - below). We can
calculate
both of them from the Snell's law.
\begin{equation}
{{\sin{\alpha}}\over{\sin{\beta_e}}} = n_{e,0} \mbox{~~~and~~~}
{{\sin{\alpha}}\over{\sin{\beta_o}}} = n_o.
\label{razm}
\end{equation}
In the expression (\ref{razm}) the angle $\alpha$ is the controlled
incident angle and $\beta_o, \beta_e$ are
the refraction angles of the ordinary and the extraordinary polarized light.
Both refraction angles are obtained from the
direction of refracted light $\gamma_o$ and $\gamma_e$ (see Fig. 3), which
can be calculated from the positions of the light dots on the screen
($x_o$ and $x_e$), the distance $l$~between the cell and the screen  and the wedge
angle $\delta$ (Fig.2) expressed in radians which is
approximatelly $d/h$ (Fig.1 a). Using the identity
\begin{equation}
{{\sin\alpha}\over{\sin
\beta}}={{\sin(\gamma\mp\delta)}\over{\beta\mp\delta}}=n
\label{lomni}
\end{equation}
the refracted beam direction $\beta$ is found from
\begin{equation}
\tan \beta = {{\delta \; \sin \alpha}\over{\sin \gamma \pm \sin
\alpha}}.
\label{kot}
\end{equation}
In the expressions (\ref{lomni},\ref{kot}) the upper sign in $\mp,\pm$
stands for the beam direction given in Fig. 3 by solid line and the
opposite sign stands for angles in the opposite direction
(dashed line - Fig. 3) .
The angle $\delta$ of the wedge cells is small ($\approx
1^\circ$) therefore $\sin \delta \approx \delta$, expressed in radians, and
$\cos\delta \approx 1$.
With the known direction $\beta$, the value of the refraction index is given by the
first part of Eq. (\ref{lomni}). Expressions (\ref{lomni}) and
(\ref{kot}) are general expressions which can be used to obtain
the
refraction index when light passes through the thin wedge sample of any, not
necessary birefringent, material and allows for evaluation of both, ordinary
and extraordinary refraction indeces.

When light passes the transverse cell the oscillating electric field has
 always one component parallel to
the long molecular axes and the other perpendicular to it. Both indeces
are independent of the incident direction as seen in Fig 1a below. The cell can be used to
demonstrate the splitting of the beam, to show the polarization of the
ordinary and extraordinary beam and to make a quantitative measurement
of the maximum value of the extraordinary refraction index and the value
of the ordinary refraction index. In adition, measurements of the dependence
of both
refractive indeces on the incident angle, although they
are constant, provide an estimation of the experimental accuracy.

In the perpendicular cell, the optical axes is perpendicular to the
glass plate and the refraction angle $\beta_e$ is equal to $\theta$ (Fig.
1b - below):
\begin{equation}
{{\sin{\alpha}}\over{\sin{\beta_e}}} = n_e(\beta_e) = n_e(\theta)
\end{equation}
where $n_e$ is calculated from Eq.~(\ref{lomni}) and (\ref{kot}).
When the incident light is perpendicular to the cell, the electric
field is perpendicular to the long molecular axis so the light
direction coincides with optical axis and the beam splitting does
not occur. But as soon as the incident light is not perpendicular
to the cell, there is a component of the electric field along the
long molecular axis and consequently the value of the
extraordinary index changes from the value of the ordinary index
(Fig 1b below). When we rotate the cell, the  beam  splits and the
single spots splits into two, demonstrating that the indeces are
not equal anymore. From the measurements of the dot positions the
behaviour of the extraordinary refractive index close to the
optical axis can be calculated.

In the parallel cell, the
optical axes is parallel to the glass and the refraction angle $\beta_e =
\frac{\pi}{2} -\theta$. The extraordinary refractive index is
therefore given by:
\begin{equation}
{{\sin{\alpha}}\over{\sin{\beta_e}}}=
{{\sin{\alpha}}\over{\sin(\frac{\pi}{2}-\theta)}} = n_e(\beta_e) =
n_e(\theta).
\end{equation}
The last of the three cells can be used to study the direction
dependency of the extraordinary index when the difference of both
indeces is close to the largest value. When incident direction of
light is varied, the extraordinary index decreases, since the
component of the oscilating electric field along the long
molecular axis decreases (Fig. 1c below). Unfortunately, to show
the decrease in the index, the evaluation of both indeces from the
measurement of the dot positions is necessary.

There are few limitations in these three experiments which have to
be considered. Although light with ordinary and extraordinary
polarization can propagate in any direction, experimentally we are
limited with the refraction of the incident light, since the light
source is outside of the birefringent material. In the presented
situation the ordinary beam direction was theoretically limited
(for the light parallel to the cell surface) to $26^\circ $ for
$n_o = 1.5$ and to
 $22^\circ$ for the extreme value of $n_{e,0}=1.76$
in the geometry of the parallel cell.
The cell size and the cell holder aditionally limit
the incident angle to the values of less than 50$^\circ$. In Fig. 4 we show
the combined results of the extraordinary index measurements in the
paralell and the perpendicular
cell and compare them to the theoretical expression (1).

To conclude, the experimental setup which enables a detailed study
of the light behavior in an uniaxial birefringent material, is
presented. In order to study the direction dependency of the
extraordinary refraction index, three different liquid crystaline
cells were designed. The transverse cell enables a demonstration
of the splitting of a nonpolarized beam into the ordinary and the
extraordinary beam. Their polarization can be shown of the extreme
value for the extraordinary refraction index as well as the value
of the ordinary refractions index can be measured. The transverse
cell can also be used to estimate the measurement accuracy. The
perpendicular cell enables the demonstration of the direction
dependecy of the extraordinary index, as well as its measurement,
when the value is close to the value of ordinary refraction index.
The parallel cell can be used for the quantitave measurement of
the extraordinary index direction dependency close to its largest
value.

The author is gratefull to Miha \v Skarabot for producing the
wedge cells, to Ludwik Kowalsky for many stimulating discussions,
to Martina \v Subic for the control set of measurements and to
Nata\v sa Vaupoti\v c for many helpfull comments during the
preparation of the manuscript.

\vspace{5mm} \noindent
$^1$E. Hecht, {\it Optics}, 3rd ed. (
Addison Wesley Longman, 1998), p. 333.\\
$^2$L. Kowalsky, private communications.\\
$^3$P. G. de Gennes and J. Prost, {\it The
Physics of Liquid Crystals}, 2nd ed., (Oxford Science
Publications, 1995), p.98.\\
$^4$D. K. Shenoy, "Measurements of
Liquid Crystal Refractive Indices", Am. J. Phys. 62, 858-859
(1994).\\
$^5$Nematic liquid crystals
with appropriate properties are commercially available by Merck.\\
$^6$M. Born and E. Wolf, {\it Principles of Optics}, 6th
ed., (Pergamon Press, 1993), p.680.
\newpage
\begin{figure}
\label{fig1} \caption{a) In the {\it parallel cell} molecular long
axes are oriented along the wedge. If the incident plane is
parallel to the longer side of the cell, the two indeces do not
change with the light direction (below). b) In the {\it
perpendicular cell} molecular long axes are oriented
perpendicularly to the cell glass. The extraordinary index
increases with the increasing incident angle (below). c) In the
{\it longitudinal cell} molecular long axes are oriented parallel
to the longer side of the cell. The extraordinary index decreases
from its maximum value with growing incident angle.}
\end{figure}
\begin{figure}
\caption{Schematic presentation of the experimental setup.}
\end{figure}
\begin{figure}
\caption{Geometry of the experiment. The light incident on the cell with the
incident angle $\alpha$, refracts in the material by an angle
$\beta$. The second incident angle is $\beta-\delta$ for the
positive $\alpha$ (solid line) and $\beta +\delta$ for the negative
$\alpha$ (dashed line). The light refracts again by an angle $\gamma$.}
\end{figure}
\begin{figure}
\label{fig4}
\caption{The extraordinary refraction index at the angle $\theta$
close to 0 measured in the perpendicular cell and at $\theta$ close to
$\pi/2$ measured in the longitudinal cell. The theoretical
direction dependence for $n_{e,0} = 1.76$ and $n_o = 1.5$ is given for a comparison.
For experiments the nematic liquid crystal E8 was used. Both indeces $n_{e,0}$
and $n_o$ were measured in the parallel cell.}
\end{figure}
\end{document}